\documentclass[a4paper]{jpconf}

\usepackage{graphicx}

\newcommand{\beq}{\begin{equation}}
\newcommand{\eeq}{\end{equation}}
\newcommand{\beqd}{\begin{displaymath}}
\newcommand{\eeqd}{\end{displaymath}}
\newcommand{\beqa}{\begin{eqnarray}}
\newcommand{\eeqa}{\end{eqnarray}}

\newcommand{\bmat}{\begin{displaymath}}
\newcommand{\emat}{\end{displaymath}}

\newcommand{\eq}[1]{Eq.~(\ref{#1})}

\begin{document}
\title{Non-linear rheology of layered systems - a phase model approach}

\author{Hajime Yoshino$^{1}$, Hiroshi Matsukawa$^{2}$, Satoshi Yukawa$^{1}$ and Hikaru Kawamura$^{1}$}

\address{$^1$Department of Earth and Space Science, Faculty of Science,
Osaka University, Toyonaka 560-0043, Japan\\
$^2$ Department of Physics and Mathematics,  Aoyana Gakuin University,
5-10-1 Fuchinobe, Sagamihara, Kanagawa, 229-8558, Japan}

\ead{yoshino@ess.sci.osaka-u.ac.jp}

\begin{abstract}
We study non-linear rheology of a simple theoretical model developed to
mimic layered systems such as lamellar structures under shear. 
In the present work we study a 2-dimensional version of the model which exhibits 
a Kosterlitz-Thouless transition in equilibrium at a critical temperature 
$T_{\rm c}$. While the system behaves as Newtonain fluid at high temperatures 
$T > T_{\rm c}$, it exhibits shear thinning at low temperatures $T < T_{\rm c}$. 
The non-linear rheology in the present
model is understood as due to motions of edge dislocations and resembles
the non-linear transport phenomena in superconductors by vortex motions.
\end{abstract}

\section{Introduction}

Frictional properties of lubricated system are often strongly influenced
by non-linear rheology of lubricants. Understanding of the physical
mechanism of the non-linear rheology is a very important basic issue 
in the science of friction and condensed matter physics in broader scope.
Examples of the lubricants include various soft matters \cite{Larson}, glasses and  granular
systems \cite{LN,GG}. 

Non-linear rheology is believed to arise from 
some combinations of elastic, plastic and viscous deformations.
However a unified physical understanding of the mechanism is lacking.
The purpose of the present work is to develop and analyze a simple 
statistical mechanical model 
which mimics layered systems, such as those with the
lamellar structure, under external shear stress. 
We demonstrate that inspite of its simplicity our model exhibits 
non-trivial rheological properties reminiscent of those in real materials.

\begin{figure}[h]
\begin{center}
\includegraphics[width=0.95\textwidth]{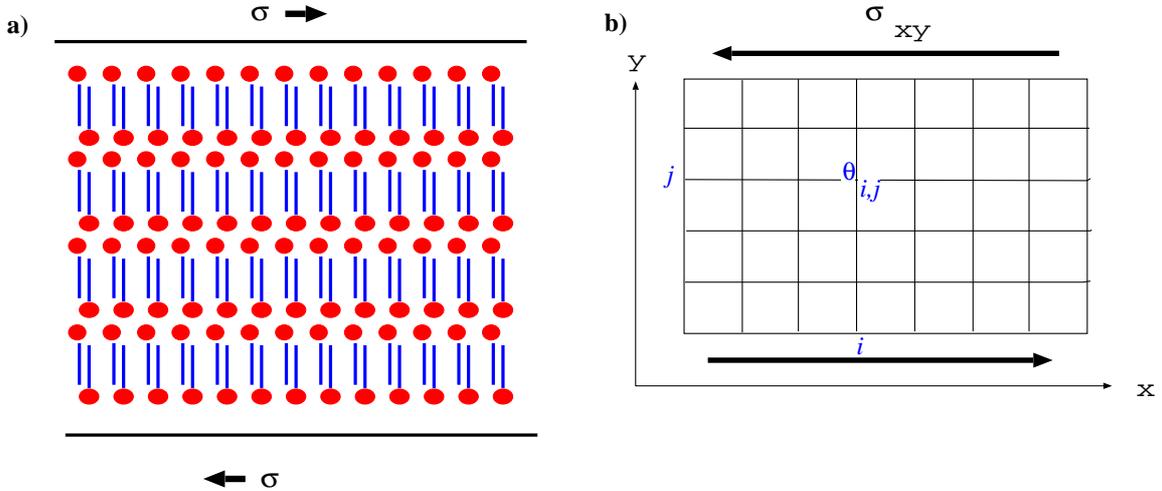}
\end{center}
\caption{A schematic picture of a lamellar structure under shear
and our lattice model.
a) Lamellar structures may be formed for example by surfactants
immersed in water. 
By applying external shear stress $\sigma$ on the top and bottom planes
the lamellar will exhibit elastic, plastic and viscous responses. b)
Our lattice model mimic such a layered system stacked into the direction
of the $y$-axis under external shear $\sigma_{xy}$ applied in the
direction of the $x$-axis. The 'phase' $\theta_{(i,j)}$
represents the displacement at lattice site $(i,j)$ into the $x$-direction.
Notice that the lattice site $(i,j)$ represents a fixed position 
with respect to the laboratory frame. 
We denote the lattice size into the direction of the $x$-axis
as $L_{x}$ and that of the $y$-axis as $L_{y}$.
}
\label{fig-model}
\end{figure}

The organization of the present paper is as the following. 
In the next section we define our model where we also sketch a close 
connection between our rheological problem 
and the transport problem in superconductors.
In sec. \ref{sec-EQ} we analyze the phase transition in our model
by a renormalization group theory and Monte Calro simulations.
Then in sec. \ref{sec-NEQ} we 
analyze the {\it flow curve}, i. e. relation between the shear-stress
and shear-rate  in our model using Langevin simulations. We analyze
the data in terms of a scaling ansatz which is analogous to that
for the current-voltage relation in superconductors. 
Finally in sec. \ref{sec-conclusions} we present our conclusions.

\section{Model}
\label{sec-model}

We consider a layered system which is an assembly of
flat elastic sheets stacked on top of each other 
as shown in Fig.~\ref{fig-model}. Interactions between the
elastic sheets consist of two parts 1) mechanical coupling which
allows  both elastic and plastic deformations and 2) viscous coupling due
to the presence of some solvents such as water. 

\subsection{Hamiltonian: a simple model with elastic and plastic deformations}

For simplicity we consider a 2-dimensional model in the present work
replacing the elastic planes by one-dimensional elastic strings. 
More specifically we consider a 2-dimensional square lattice model 
of size $L_{x} \times L_{y}$ shown in Fig.~\ref{fig-model} b). 
Extension to a three-dimensional model is straightforward.

We define dimensionless 'phase' variables $\theta_{(i,j)}$'s 
on lattice sites $(i,j)$'s assuming that spatial profile of the density
is specified as $\rho(i,j)=\rho_{0}+\rho_{1}\cos(\theta_{(i,j)})$ with
$\rho_{0}$ and $\rho_{1}$ being certain constants. 
The interactions between the phase variables are described 
by the following Hamiltonian,
\beq
H=\frac{J}{2}\sum_{\langle i,j \rangle} (\theta_{(i+1,j)}-\theta_{(i,j)})^{2}
-J'\sum_{(i,j)} \cos(\theta_{(i,j)}-\theta_{(i,j)+1}).
\label{eq-hamiltonian}
\eeq
Here $J$ and $J'$ denote the strength of the interactions. The 1st term on the
r. h. s represents the elastic couplings within each elastic layers 
and the 2nd term 
is a sinusoidal coupling which allows elastic and plastic deformations
between adjacent elastic layers. 
Although the strength of the two couplings should
be different in general, we choose both of them to be $J=J'=1$ for
our convenience. We note that the effective Hamiltonian \eq{eq-hamiltonian}
is a much simplified one as compared with more realistic
expressions which take into account {\it splay distortions}
of the layers\cite{CL}.

\begin{figure}[h]
\begin{center}
\includegraphics[width=0.95\textwidth]{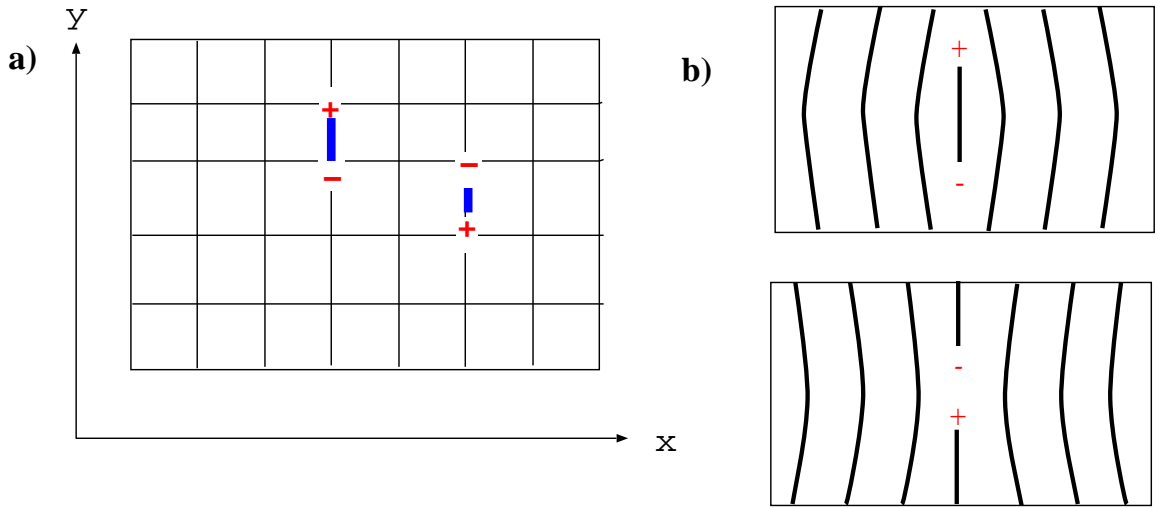}
\end{center}
\caption{Elementary dislocation-dipoles. 
If the difference $\Delta \theta$ between the two phase variables 
associated with a interaction bond oriented in the $y$-direction becomes
greater in magnitude than $\pi$, the bond {\it slips}. In an analogy with 
the electromagnetism, such an an elementary plastic deformation can be regarded
as creation of a {\it dislocation-dipole} of unit length $l_{0}$
with $\pm 1$ changes on both ends as shown in a). Note that 
the dipoles are parallel to the $y$-axis.
The two possible orientations of the dipoles correspond to the sign
of the phase slips $\Delta \theta$. The latters correspond to the two types
of edge dislocations created in the crystal as shown in b)
where the region of high densities are marked by lines.
}
\label{fig-dipole}
\end{figure}

Apparently the ground state is given by a spatially uniform $\theta$
which corresponds to a crystalline structure.
Low energy excitations from the ground state are 1) Goldstone modes: smooth
spacial variation of $\theta$ and  2) plastic deformations due to
dislocation-dipoles  as shown in Fig.~\ref{fig-dipole}.
An elementary dislocation-dipole consists of a pair of $\pm 1$ 'charges'
on its two ends separated by a lattice spacing $l_{0}$.
In a representation of the configuration of the phase variables
$\theta_{(i,j)}$ in terms of 
unit vectors $(\cos \theta_{(i,j)},\sin \theta_{(i,j)})$
as the XY spins, the $\pm 1$ charges appear as vorticies and anti-vorticies.

Periodic boundary condition is imposed in the $x$-direction while
phase variables at the top/bottom layers ($j=L_{y},1$) 
are regarded as to represent 'walls'. We specify their properties 
shortly later.

\subsection{Dynamics}

We model the dynamics by the following Langevin equation for the
phase variables $\theta_{(i,j)}$  and their velocity $v_{(i,j)}$,
\beqa
&& \frac{d \theta_{(i,j)}}{dt}=v_{(i,j)}  \nonumber \\
&& m\frac{d v_{(i,j)}}{dt}=-\eta_{0} \sum_{j'=j \pm 1}(v_{(i,j)}-v_{(i,j')})
 -\frac{\partial H}{\partial \theta_{(i,j)}}+\xi_{(i,j)}(t)
\label{eq-langevin}
\eeqa
where $m$ is the effective mass of the phase variables which we choose
to be $m=1$ for our convenience. The 1st term on the r.~h.~s. 
of the 2nd equation represents viscous couplings between adjacent
elastic layers.  We choose the {\it bare} viscosity to be $\eta_{0}=1$ 
for our convenience. The 2nd terms is due to the mechanical
couplings and the last term is the thermal noise.

The thermal noise $\sigma_{(i,j)}(t)$ follows the Gaussian distribution
with zero mean and the following spatio-temporal correlations,
\beq
\langle \xi_{(i,j)}(t) \xi_{(i',j')}(t') \rangle =
2 \eta_{0} k_{\rm B} T 
\delta (t-t')\delta_{i,i'}(\delta_{j,j'}-\delta_{j,j'-1}-\delta_{j,j'+1}). 
\eeq
Note that the correlation between the thermal noise at neighbouring layers
is needed due to the viscous (dissipative) copling between them.
Here $k_{\rm B}$ is the Boltzmann's constant which we put to be $1$
for our convenience.  In the following we use
$\langle \ldots \rangle$ for thermal averages.

The phase variables $\theta_{(i,j)}$ on the top ($j=L_{y}$)
and bottom ($j=1$) layers are regarded as to belong to
'rigid walls' which are driven into the opposite directions.
More precisely we model the walls as
$\theta_{(i,L_{y})}=\theta^{I}$  and $\theta_{(i,1)}=\theta^{II}$ 
which are expressed via Fourier series,
\beq
\theta_{i}^{I(,II)}= \sum_{n=1}^{L_{x}}
\frac{a_{n}^{I,(II)}}{\sqrt{L_{x}}}
\cos \left [\frac{2n\pi}{L_{x}}\left(i-\frac{x^{I(,II)}_{\rm CM}}{l_{0}}\right)
+\phi_{n}^{I,(II)} \right].
\eeq 
Here $x_{\rm CM}$ represents the center of mass position of the wall
and $l_{0}$ is the lattice spacing. We choose $l_{0}=1$ in the following. 
To mimic a 'rough wall' we choose random values for $a_{n}$s 
drawn from a Gaussian distribution
of zero mean and variance $1$ while we choose random values for 
$\phi_{n}$s from a uniform distribution between $0$ and $2 \pi$. 
Obviously a 'regular wall' can be selected as well by 
choosing a certain values for $a_{n}$ and $\phi_{n}$.

In the present work we drive the walls at constant velocities by enforcing, 
\beq
x^{I(,II)}_{\rm CM}=\pm v_{\rm wall}(t)
\eeq
where $\pm$ is for $I/II$. We define the {\it apparent} shear-rate $\dot{\gamma}$
as,
\beq
\dot{\gamma}=\frac{v_{\rm wall}}{L_{y}}.
\eeq

\subsection{Relation to the transport problem in superconductors}

Here let us mention briefly a remarkable connection between our
rheological problem and the transport problem in superconductors \cite{FFH}.

Apparently a very important issue in the problem of superconductivity is the
macroscopic transport property: how the Ohmic resistance in the normal 
phase disappear as the superconductivity sets-in.
Essential macroscopic properties of the superconductivity are determined
by ordering of the {\it phase} of its order parameter. The 
phase can be identified with the phase variable $\theta$ in our model
and the effective hamiltonian can be given as  ours \eq{eq-hamiltonian}
but with the elastic coupling in the $x$-direction replaced by
a sinusoidal one, i.e. the usual isotropic XY model. 
Then our two-dimensional model correspond to a superconducting film \cite{WGI}.
Shortly later we discuss similarity and differences between 
the equilibrium properties of our model and the usual XY model.

A standard model to study macroscopic transport properties in
superconductors are the so called resistively-shunted-junction (RSJ)
model (see for example \cite{MJP}) in which the couping $J$
in \eq{eq-hamiltonian} is regarded as the strength of the Josephson coupling
between superconducting grains. The mass $m$ and the bare viscosity 
$\eta_{0}$ in our model \eq{eq-langevin} are regarded as the capacitance
and the inverse of the so called shunted-resistance between
superconducting grains respectively. 
External forces $\pm I$ are applied on the top/bottom layers
to mimic in-coming and out-going external electric currents.
The latters correspond to 
nothing but the external shear stresses $\sigma_{xy}$ in our problem.
One then measures the voltage drop $V$ induced in the system which
corresponds to the shear rate $\dot{\gamma}$ in our problem.
Thus the $I-V$ (current vs. voltage) characteristic in
the superconductors corresponds to the $\sigma_{xy}-\dot{\gamma}$ 
(shear-stress vs. shear-rate) relation
which is called as {\it flow curve} in rheology.

\section{Equilibrium Phase transition}
\label{sec-EQ}

Here let us discuss some essential features of
the phase transition in the present model
which will provide us a useful basis to analyze the rheology of the model. 

First note that our model 
given by the hamiltonian \eq{eq-hamiltonian} is similar to the ferromagnetic
XY spin model. The difference is that the couplings in the $x$-direction
is elastic in our model while the couplings are sinusoidal in both $x$
and $y$ directions in the XY model. In the XY models an elementary plastic 
deformation is creation of a pair of $\pm 1$
charges, which need not to form the specific 
type of the dipoles shown in Fig.~\ref{fig-dipole}. In superconductors
the $\pm 1$ charges correspond to the quantized 
vorticies and anti-vorticies.

It is well known that the XY model exhibits the Kosterlitz-Thouless (KT)
transition at a finite critical temperature \cite{KT}. We now wish to clarify whether
our model, which is extremely anisotropic, also exhibits a similar 
phase transition. To this end we set-up a renormalization group theory.

First by taking a continuous limit we obtain 
an effective model of a scalar field
$\psi(x,y)$ in the 2-dimensional space, whose partition function is given as,
\beq
Z = \int {\cal D}\psi¡¡e^{-{\cal S}} 
\qquad {\cal S}=
\int d^{2}{\bf x} \left[
\frac{\beta J}{2} (\nabla \psi)^{2}
- \mu_{\rm d} \cos \left ( 2 \pi (\beta J) l_{0} \frac{\partial
\psi}{\partial y} 
\right) \right].
\eeq
where $\beta=1/k_{\rm B}T$ is the inverse temperature, $l_{0}$
is a unit length in the lattice model and
$\mu_{\rm d}$ is the fugacity of a dislocation-dipole.
Although the above model resembles the the sine-Gordon model
\cite{Kogut} which can be obtained by taking a continuous limit of the XY model,
the argument of the cosine function in the 2nd term of the action ${\cal S}$ 
is the derivative $l_{0}\frac{\partial \psi}{\partial y}$ while it is
$\psi$ in the usual sine-Gordon model.

\begin{figure}[h]
\begin{center}
\includegraphics[width=0.45\textwidth]{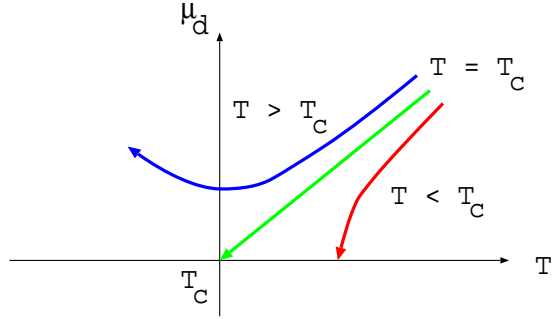}
\end{center}
\caption{Renormalization group flow of the fugacity $\mu_{\rm d}$ of the
dislocation-dipoles vs temperature $T$.
For the 'bare' temperatures $T \leq T_{\rm c}$, the fugacity is renormalized
to $0$ meaning that after coarsegraining 
the entire system can be regarded as 
an dislocation-free {\it elastic} body at some effective
 (renormalized) temperature. 
At $T > T_{\rm c}$, the flow of $\mu_{\rm d}$ 
exhibits an upturn meaning that the
dislocations are relevant at high temperatures where the elastic layers
flow smoothly on top of each other.
}
\label{fig-dipole-rg}
\end{figure}

We analyze the renormalization group (RG) flow of the temperature $T$ and 
the fugacity of the dislocation-dipoles.
The RG flow represented in the $T$-$\mu_{d}$ plane
is shown in Fig.~\ref{fig-dipole-rg}. Remarkably it reveals a
KT transition  \cite{KT,Kogut} at a critical temperature $T_{\rm c}$
as in the usual sine-Gordon model in spite of the strong anisotropy 
in our model. 

In the RG analysis of our model we are forced to follow not only 
the flow of the temperature $T$ and the fugacity $\mu_{d}$
but also the ratio between the correlation length
$\xi_{\parallel}$ in the $x$-direction and $\xi_{\perp}$ in the
$y$-direction. We find the ratio behaves non-trivially as,
\beq
\ln \frac{\xi_{\parallel}}{\xi_{\perp}} \propto 
\left (\frac{T-T_{\rm c}}{J}\right)^{3/2} \qquad T > T_{\rm c}
\eeq
in the high temperature phase. It means that the system
tends to order more strongly within each elastic layer than between
different elastic layers as expected. Note however that ratio converges
to just a constant as $T \to T^{+}_{\rm c}$ implying that
the anisotropy does not change the universality.
As in the usual KT transition, the correlation lengths themselves diverge
exponentially fast as $T \to T^{+}_{\rm c}$,
\beq
\xi_{\perp}  \sim l_{0}\exp \left[ A \left(\frac{T-T_{\rm c}}{J}\right)^{-1/2} \right] \qquad T > T_{\rm c}.
\label{eq-xi-kt}
\eeq
where $A$ is a numerical constant.
In the whole low temperature phase $T < T_{\rm c}$ the correlation length
$\xi$ of the fluctuation 
remains $\infty$, i.e. the long-range order is absent and the
system remains critical.

For later analyses we need to know the precise value of the 
critical temperature $T_{\rm c}$ of the original
lattice model. To find out $T_{\rm c}$, we analyzed 
the relaxation of an auto-correlation function 
\beq
C(t)=(1/N)\sum_{(i,j)}\langle
\cos(\theta_{(i,j)}(t)-\theta_{(i,j)}(0))\rangle 
\eeq
starting from a random initial condition at time $t=0$
\footnote{Random initial configurations are realized by choosing $\theta_{(i,j)}(0)$'s 
out of a uniform distribution between $0$ and $2\pi$.}.
In the cases of 2nd order phase transitions, including the KT transition,
such an auto-correlation function is expected to decay as
\beq
C(t) \propto t^{-\lambda}e^{-t/\tau} \qquad T > T_{\rm c}, 
\eeq 
in the high temperature phase $T \geq T_{\rm c}$.
The relaxation time $\tau$ is related the correlation length $\xi$ via
\beq
\tau \sim \tau_{0} (\xi/l_{0})^{z}
\label{eq-tau-xi}
\eeq
where $z (> 0)$ is the dynamical critical exponent and 
$\tau_{0}$ is the microscopic time scale associated with
the microscopic length scale $l1_{0}$.
Since the correlation length
$\xi$ diverges in the limit $T \to T_{\rm c}^{+}$, $\tau$ also diverges
as well.
Then right at the critical temperature $T_{\rm c}$, $C(t)$ exhibits
a purely power law decay $t^{-\lambda}$ with some exponent $\lambda (>0)$.

In practice we  used the heat-bath Monte Calro (MC)
method and simulated relaxations in large systems of sizes $320\times 320$
and $640 \times 640$ by which we could observe $C(t)$ without 
appreciable finite size effects up to $t=10^6$ MC steps (MCS).
By analyzing $C(t)$ at various temperature we found 
$T_{\rm c}/J \simeq 1.15 \pm 0.03$  and $\lambda \simeq 0.067 \pm 0.002$.
The latter value of $\lambda$ agrees with that found in the
2-dimensional XY model \cite{OOI} suggesting again that the present model
belongs to the same universality class as the 2-dimensional XY model.

\section{Non-linear rheology}
\label{sec-NEQ}

Now let us discuss the non-linear rheology of the present model
based on our results obtained by numerical simulations of 
the Langevin eq. \eq{eq-langevin} under constant external shear rate
$\dot{\gamma}$.

\begin{figure}[t]
\begin{center}
\includegraphics[width=0.95\textwidth]{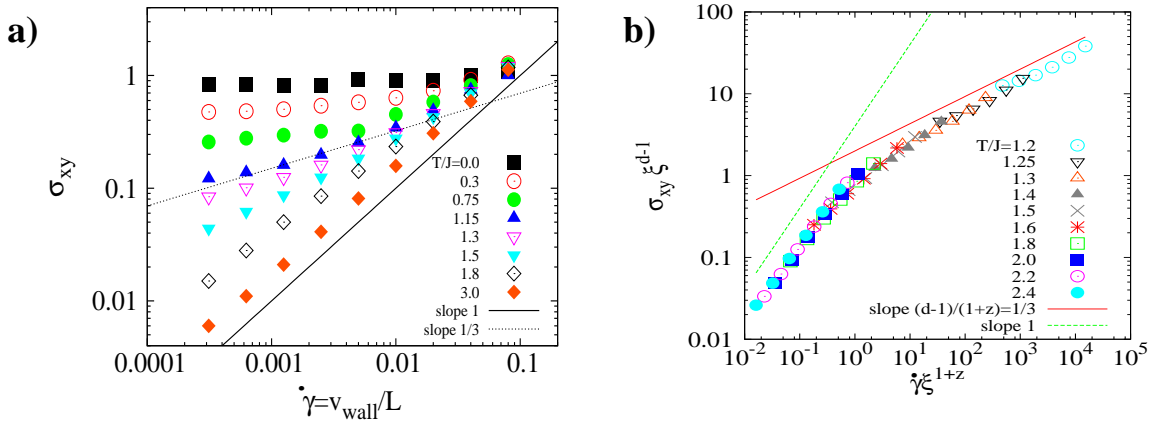}
\end{center}
\caption{The flow curve and its scaling plot.
a) Shear-stress $\sigma_{xy}$ is plotted
versus the shear-rate $\dot{\gamma}$. We used systems of sizes
$40 \times 40$. By comparing with the data of $60 \times 60$
we found finite size effects are not significant within the
range of shear-rates used in the figure.
b) The scaling plot for the data at $T> T_{\rm c}$
at shear rates $\dot{\gamma} < 0.01$ using the scaling ansatz
given by \eq{eq-scaling-ansatz}. Here we used
$d=2$, $z=2$ and the correlation length
given by \eq{eq-xi-kt} with $T_{\rm c}=1.15$ determined in sec. \ref{sec-EQ}.
The only adjustable parameter used in this scaling plot is the
numerical factor $A$ in \eq{eq-xi-kt} which we choose as
$A=1.5$.}
\label{fig-flow}
\end{figure}

\subsection{Flow curve}
\label{subsec-flow-curve}

We first examine the flow curve, i.e. the relation between the 
shear-rate $\dot{\gamma}$ and shear-stress $\sigma_{xy}$
in the stationary state. To this end we have performed simulations under constant
shear-rate $\dot{\gamma}=v_{\rm wall}/L_{y}$ and
evaluated the resultant shear-stress $\sigma_{xy}$ between adjacent layers,
\beq
\sigma_{xy}= \frac{1}{L_{x}} \sum_{i=1}^{L_x}
 \left \{
J' \sin(\theta_{(i,j+1)}-\theta_{(i,j)}) 
+ \eta_{0} (v_{(i,j+1)}-v_{(i,j)}) \right\}.
\eeq
In Fig.~\ref{fig-flow} a) we show the data of the shear-stress
$\sigma_{xy}$ measured under various shear-rates $\dot{\gamma}$
and at various temperatures $T/J$ in the double logarithmic plot.

Except for very high shear rate region $\dot{\gamma} \gg 0.01$
where the contribution of the {\it bare} viscous coupling parametrized  
by the $\eta_{0}$ becomes dominant, the shape of the flow curve is non-trivial.

The relation between $\dot{\gamma}$ and $\sigma_{xy}$ 
clearly exhibits the {\it Newtonian fluid} behaviour 
$\sigma_{xy}=\eta \dot{\gamma}$
at high temperatures $T > T_{\rm c}$ under low enough shear rates
$\dot{\gamma}$. Apparently the viscosity $\eta$ 
increase as the temperature is lowered toward $T_{\rm c}$ 

At the same time, {\it shear-thinning}\footnote{The effective shear
viscosity behaves as
$\eta_{\rm eff}=\sigma_{xy}/\dot{\gamma} \sim \gamma^{-\alpha}$
meaning that the system flows more easily at higher shear rates.
Such a behaviour is observed in a variety of systems including
various soft matters, glasses and granular systems.
}
behaviour $\sigma_{xy} \propto \dot{\gamma}^{1-\alpha}$ emerges with some
exponent $\alpha  \sim 1/3$  under higher shear rates
except in very high shear rate region $\dot{\gamma} > 0.01$.
Remarkably such a power-law region extends to lower shear rates as the 
temperature is lowered. 
Right at $T=T_{\rm c}$ it appears that the power law behaviour 
dominates the entire range of the shear rate except the 
very high shear rate region $\dot{\gamma} > 0.01$. 

In the low temperature phase $T < T_{\rm c}$,
some finite {\it yield stress} $\lim_{\dot{\gamma} \to 0}\sigma \neq 0$
would emerge if some long range order is established.
However, long range order is absent in the present 2-dimensional
model as noted before in sec.~\ref{sec-EQ}
and the system remains critical in the whole temperature range
$T < T_{\rm c}$.
Indeed the power law behaviour 
$\sigma_{xy} \propto (\dot{\gamma})^{1-\alpha(T)}$ continues down to lower
temperatures with a temperature dependent exponent $\alpha(T)$ 
which increases up to $1$ as $T \to 0$ 
\footnote{We note however that if we plot the flow curve not in the
double logarithmic plot as in Fig.~\ref{fig-flow} a) 
but in a linear plot (not shown),
we would be tempted to conclude that $\lim_{\dot{\gamma} \to 0}\sigma \neq 0$
below $T_{\rm c}$ because of the {\it very slow} decrease of $\sigma_{xy}$
as $\dot{\gamma}$ is decreased.}.

\subsection{Scaling law for the flow curve}

\begin{figure}[t]
\begin{center}
\includegraphics[width=0.45\textwidth]{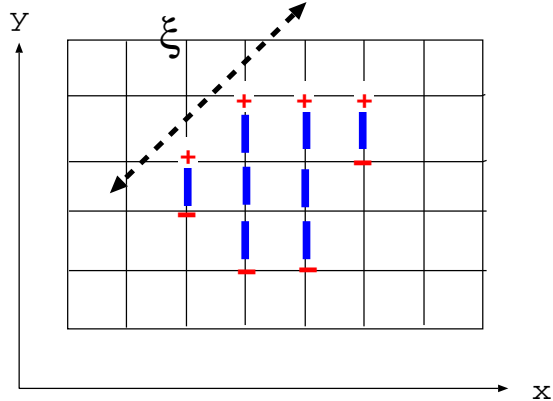}
\end{center}
\caption{A schematic picture of a collective excitation
of dislocation-dipoles. By creating elementary dislocation-dipoles
coherently in a closed volume,
the interior of the volume exhibits a uniform phase slip $2 \pi$
with respect to the rest of the system.
}
\label{fig-collective-dipoles}
\end{figure}

The basic features of the flow curve discussed above suggest that there might
be a scaling law which explains the flow curve in a unified manner. 
In the following we first rephrase a scaling ansatz\cite{FFH}, which is
originally proposed for the non-linear current-voltage relation in
superconductors, within our context of rheology.
Then we examine the validity of the scaling ansatz based on our data.
In the following we disregard the trivial contribution from the bare 
viscous force which can be neglected closer to $T_{\rm c}$ under low enough shear-rates.

Let us consider a cluster of volume $\xi^{d}$ in which the phase
variables are strongly correlated with each other. Presumably the life time
of such a cluster is of the order of the relaxation time 
$\tau\sim \tau_{0}(\xi/l_{0})^{z}$ as given in \eq{eq-tau-xi}.
Under a given shear rate $\dot{\gamma}$, the total phase difference 
across the cluster will become 
$\Delta \theta = \dot{\gamma} \tau \xi$  within the life time $\tau$.
The cluster will make a phase slip if $\Delta \theta \sim 2\pi$, i.e.
\beq
\dot{\gamma} \tau_{0} \left(\frac{\xi}{l_{0}}\right)^{1+z} \sim 2\pi.
\eeq
Such a phase slip event may be viewed as a collective excitation of
dislocation-dipoles within a volume $\xi^{d}$ 
as shown schematically in Fig.~\ref{fig-collective-dipoles}. 
Preseumably such an event takes place as a thermally activated process.
The energy barrier associated with it will be
of order $l_{0}\sigma_{xy} \xi^{d-1}$. Thus we expect that 
the probability for the event is a function of $l_{0}\sigma_{xy} \xi^{d-1}/k_{B}T$.

Based on the above observations we propose the following scaling ansatz,
\beq
\dot{\gamma}= \tau^{-1}_{0} \left(\frac{\xi}{l_{0}}\right)^{-1-z}
f_{\pm} \left( \frac{l_{0}\sigma_{xy} \xi^{d-1}}{k_{\rm B}T} \right).
\label{eq-scaling-ansatz}
\eeq
where the scaling functions $f_{+}$ and $f_{-}$ are for
the high temperature phase $T>T_{\rm c}$ and low temperature phase
$T < T_{\rm c}$ respectively.

\subsubsection{High temperature phase}

In the high temperature phase $T> T_{\rm c}$, it is natural to expect
that the system behaves as a Newtonian fluid,
correspnding to the {\it Ohmic resistivity}, 
as far as the correlation 
length $\xi$ remains finite. Right at $T=T_{\rm c}$, $\xi$ diverges
so that dependence on $\xi$ must be eliminated. Then a natural scaling ansatz
for $f_{+}(x)$ is,
\beq
f_{+}(x) \sim \left \{
\begin{array}{cc}
x & x \ll 1 \nonumber \\
x^{\frac{1+z}{d-1}} &  x \gg 1
\end{array}
\right.
\eeq
It means that the macroscopic shear viscosity $\eta=\sigma/\dot{\gamma}$ 
exhibits an anomalous scaling close to $T_{\rm c}$,
\beq
\eta \propto \xi^{d-2-z}
\eeq
and that the flow curve exhibits a purely power law, {\it shear-thinning}
behaviour right at $T=T_{\rm c}$,
\beq
\sigma \propto \dot{\gamma}^{1-\alpha} 
\eeq
with the shear-thinning exponent,
\beq
\alpha=\frac{2-d+z}{1+z}.
\label{eq-alpha}
\eeq

At the lower critical dimension $d=2$ and upper critical 
dimension $d=4$, the dynamical exponent is $z=2$ \cite{FFH}. Thus
one finds $\alpha=2/3$ for $d=2$ and $\alpha=0$ for $4$ respectively.
The latter implies shear-thinning is absent in $d > 4$ where
mean-field theories hold.

We have indeed observed $\alpha \sim 2/3$ at $T_{\rm c}$ as we noted
before (See Fig.~\ref{fig-flow} a)). Correspondingly, in a
superconducting film $V \sim I^{3}$ behaviour has been observed
experimentally \cite{WGI}.

Now let us examine the scaling ansatz using our data of the shear-stress
$\sigma_{xy}$ obtained at various shear rates $\dot{\gamma}$
and various temperatures $T > T_{\rm c}$. As shown in Fig.~\ref{fig-flow} b)
the scaling ansatz explains very well the crossover from the Newtonian
fluid regime to the shear-thinning regime.

\subsubsection{Low temperature phase}

In the low temperature phase, just below $T_{\rm c}$, the system should
exhibit the shear-thinning again. On the other hand, at lower 
temperatures the Arrhenius law will hold
since the plastic deformations take as thermally activated
processes. Thus a natural scaling ansatz for $f_{-}(x)$ is,
\beq
f_{-}(x) \sim \left \{
\begin{array}{cc}
e^{-x} & x \gg 1 \nonumber \\
x^{\frac{1+z}{d-1}} &  x \ll 1
\end{array}
\right.
\eeq
It means that {\it linear} viscosity, corresponding to the Ohmic
resistance,  vanishes in the
$\dot{\gamma} \to 0$ limit. However note that
the {\it yield stress}, 
corresponding to the {\it critical currents} in superconductor,
defined by {\it strictly}
taking the limit $\lim_{\dot{\gamma}\to 0}\sigma_{xy}$ 
is {\it zero} at {\it any} finite temperatures 
due to the presence of thermally activated plastic deformations,
sometimes called as {\it creep}, even in crystalline systems.

In the case of the present 2-dimensional model which exhibit the KT
transition, the whole temperature range $T < T_{\rm c}$ 
is critical in the sense that the correlation length $\xi$ of the
fluctuation remains $\infty$.
Thus the critical behaviour 
$f_{-}(x) \sim x^{-\frac{d-2-z(T)}{d-1}} $
will persist within the low temperature phase with
some temperature dependent dynamical exponent\cite{MJP} $z(T)$ which decreases
down to $0$ as $T \to 0$, i. e. the system
exhibits shear-thinning behaviour in the whole low temperature phase
with the temperature dependent 
shear-thinning exponent $\alpha(T)=z(T)/(1+z(T))$ (See \eq{eq-alpha}).

In our data shown in Fig.~\ref{fig-flow} a), the data at $T < T_{\rm c}$
indeed exhibit power law behaviour 
$\sigma_{xy} \propto (\dot{\gamma})^{1-\alpha(T)}$
with temperature dependent exponent
$1-\alpha(T)$ which decreases as the temperature $T$ is lowered.

\section{Conclusions}
\label{sec-conclusions}

In the present paper we analyzed non-linear rheology in
a simple theoretical model which mimics layered systems
such as those with the lamellar structures under shear. More specifically
we analyzed in detail a 2-dimensional model which we found to exhibit
a Kosterlitz-Thouless transition at a finite temperature $T_{\rm c}$.
The flow curve exhibits shear-thinning behaviour below $T_{\rm c}$.
The flow curve follows very well a scaling ansatz which we obtained
by translating the scaling ansatz for the non-linear transport in
superconductors to that for our rheological problem.
 
We wish to report more details of the present work together with 
some analysis on other features of the non-linear rheology in our system,
such as an apparent increase of the viscosity as the thickness 
$L_{y}$ is made smaller than the correlation length $\xi_{\perp}$, 
shear-banding and stick-slip motions, elsewhere.

\section*{References}

\bibliographystyle{unsrt}
\bibliography{refs}

\end{document}